\documentclass[journal=jacsat,manuscript=article]{achemso}

\usepackage{chemformula} % Formula subscripts using \ch{}
\usepackage[T1]{fontenc} % Use modern font encodings
\usepackage{subcaption}
\usepackage{siunitx}
\usepackage{epstopdf}

\author{Yubin Park}
\email{yubinpark@stanford.edu}
\affiliation[Stanford EE]
{Department of Electrical Engineering, Stanford University, Stanford, California 94305, USA}
\author{Viktar S. Asadchy}
\author{Bo Zhao}
\affiliation[Stanford Ginzton]
{Ginzton Laboratory, Stanford University, Stanford, California 94305, USA}
\author{Cheng Guo}
\author{Jiahui Wang}
\affiliation[Stanford AP]
{Department of Applied Physics, Stanford University, Stanford, California 94305, USA}
\author{Shanhui Fan}
\email{shanhui@stanford.edu}
\affiliation[Stanford EE]
{Department of Electrical Engineering, Stanford University, Stanford, California 94305, USA}
\alsoaffiliation[Stanford Ginzton]
{Ginzton Laboratory, Stanford University, Stanford, California 94305, USA}

\title[An \textsf{achemso} demo]
  {Violating Kirchhoff's Law of Thermal Radiation in Semitransparent Structures}

\abbreviations{CMT, BIG}

\begin{document}


\begin{abstract}
  Kirchhoff's law of thermal radiation imposes a constraint on photon-based energy harvesting processes since part of the incident energy flux is inevitably emitted back to the source.
  By breaking the reciprocity of the system, it is possible to overcome this restriction and improve the efficiency of  energy harvesting.
  Here, we design and analyze a semitransparent emitter that fully absorbs normally incident energy from a given direction with zero backward and unity forward emissivity.
  The nearly ideal performance with wavelength-scale thickness is achieved due to the magneto-optical effect and the guided-mode resonance engineered in the emitter structure.
  We derive the general requirements for the nonreciprocal emitter using the temporal coupled mode theory and the symmetry considerations.
  Finally, we provide a realistic emitter design based on a photonic crystal slab consisting of a magnetic Weyl semimetal and silicon.
\end{abstract}


\section{Introduction}
\begin{figure}
  \includegraphics[width=0.95\columnwidth]{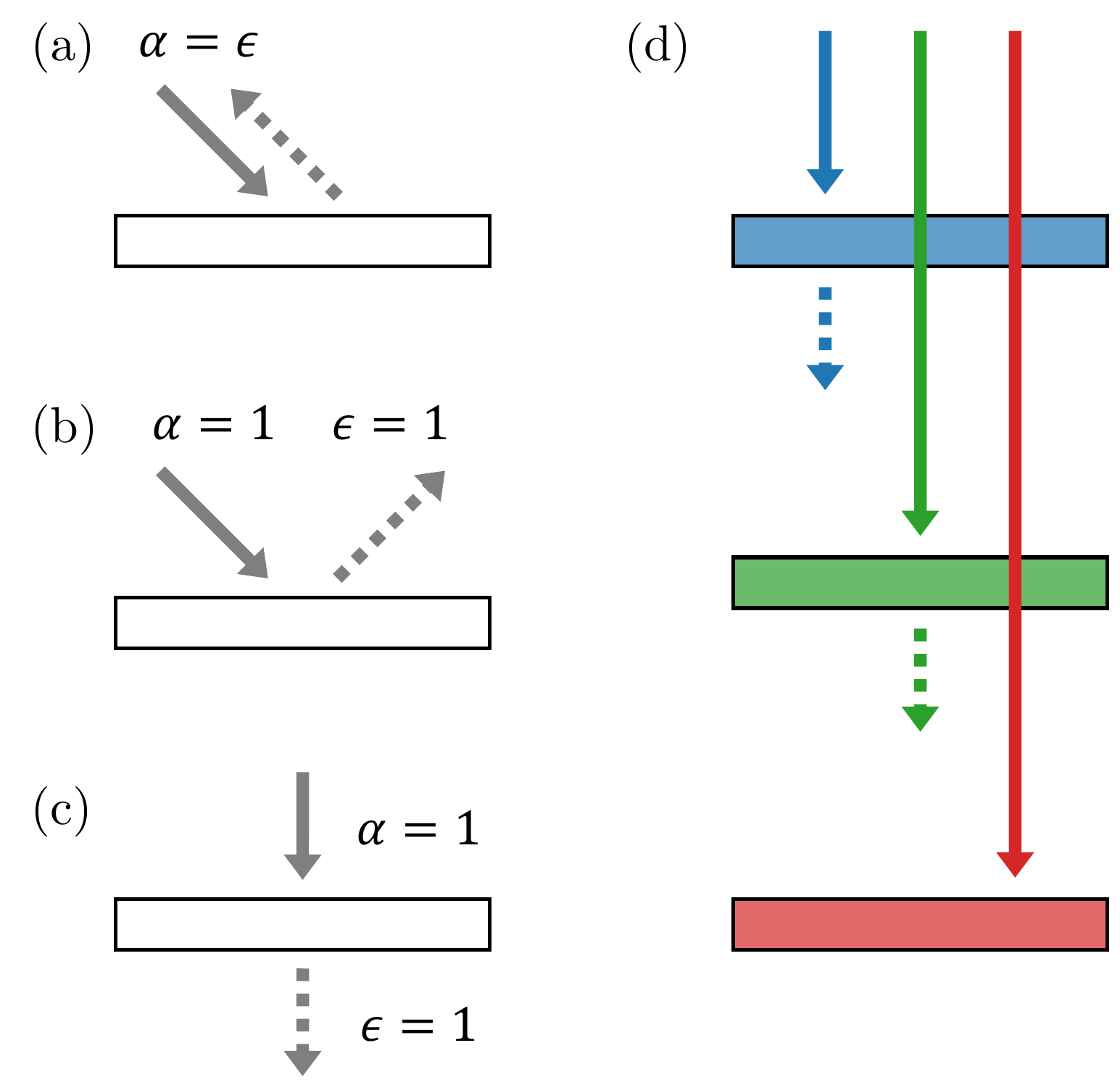}
  \caption{ (a--c) Illustrations of the energy flow in various thermal emitter systems: (a)~reciprocal case, (b)~nonreciprocal opaque case in reflective configuration, (c)~nonreciprocal semitransparent case in transmissive configuration. The arrows define the directions of energy exchange between the emitter and the environment for a given frequency (solid arrows: absorption, dotted arrows: emission). Here, $\alpha$ and $\epsilon$ represent the absorptivity and emissivity of the emitter, respectively.
  (d)~Illustration of a multi-junction solar cell with the desired nonreciprocal functionality. Each box represents a cell with a particular bandgap. The bandgap of each cell decreases from the top to bottom as expressed with blue, green, and red colors. Each colored arrow indicates the light at the corresponding frequency of the cell bandgap.}
  \label{fgr:intro}
\end{figure}

  Kirchhoff’s law of thermal radiation states that emissivity of a body has to be equal to its absorptivity for a given angle-polarization and frequency of thermal electromagnetic radiation~\cite{kirchhoff_ueber_1860, planck_theory_1914, howell_thermal_2021, bergman_fundamentals_2017}.
  Due to this balance between emission and absorption, whenever an object absorbs light, it must emit energy back to the original source, as illustrated in Fig.~\ref{fgr:intro}a.
  Processes of energy harvesting from thermal radiation inevitably suffer from this loss.
  Thus, overcoming this restriction by violating the Kirchhoff's law is of fundamental importance in order to provide opportunities for higher energy harvesting efficiency.\cite{green_time-asymmetric_2012}
  
  The Kirchhoff's law is generally obeyed by any system that satisfies the Lorentz reciprocity.~\cite{landau_electrodynamics_1984, han_theory_2009}
  Once the reciprocity of a system is broken, this law does not hold anymore and it is possible to achieve a nonreciprocal thermal emitter which exhibits a contrast between the emissivity and the absorptivity.~\cite{ries_complete_1983, snyder_thermodynamic_1998, zhang_validity_2020, khandekar_new_2020}
  Employing the magneto-optical effect is a widely used method to design a nonreciprocal system, and there have been several preceding works exploiting this effect to construct nonreciprocal thermal emitters.~\cite{zhu_near-complete_2014, zhao_near-complete_2019, wang_nonreciprocal_2017, wang_nonreciprocal_2018, pajovic_intrinsic_2020}
  \citeauthor{zhu_near-complete_2014} proposed a specific design for a thermal emitter using magneto-optical photonic crystals in Ref.~\citenum{zhu_near-complete_2014}, and \citeauthor{zhao_near-complete_2019} developed this work further by greatly reducing the required magnetic field in the structure in Ref.~\citenum{zhao_near-complete_2019}.
  In addition, various materials have been studied for the purpose of achieving nonreciprocal thermal emitters.~\cite{wang_nonreciprocal_2017, wang_nonreciprocal_2018, pajovic_intrinsic_2020}
  All these  emitters are opaque and suitable only for reflection-based nonreciprocal systems~\cite{green_time-asymmetric_2012} (see Fig.~\ref{fgr:intro}b).
  However, for certain applications it is preferable to have nonreciprocal emission in a transmission-based configuration, as depicted in Fig.~\ref{fgr:intro}c. 
  For example, in multi-junction solar cells, the p-n junctions with higher bandgaps are arranged towards the top of the solar cell.
  In this case, the emission from the junctions towards the sun represents an intrinsic photon loss mechanism.
  This loss channel can be avoided if the junctions are made nonreciprocal such that the emitted photons from the upper junctions can be absorbed by the lower-bandgap junctions below to contribute to electrical energy generation (see Fig.~\ref{fgr:intro}d).
  
  In this paper, we analyze and propose a semitransparent photonic crystal slab that can achieve the above mentioned functionality, having the total thickness of the order of the operational wavelength.
  The emitter operates for normal illumination and  provides a near-unity difference between the absorptivity and emissivity in the direction of the source as illustrated in Fig.~\ref{fgr:intro}c.
  
  The paper is organized as follows.
  First, using the temporal coupled mode theory (CMT), we  derive general theoretical conditions required for the violation of Kirchhoff's law in a semitransparent slab for normal incidence.
  Next, we determine the symmetry of the photonic crystal slab which allows us to obtain the desired thermal performance and provide a possible unit-cell topology based on the guided-mode resonance.
  This part is followed by the full-wave simulation results of the structure and the qualitative analytical description of its response.
  Finally, we provide a feasible realization of the photonic crystal slab  based on a magnetic Weyl semimetal.

\section{Theory}
\subsection{Energy balance in thermal equilibrium}
\begin{figure}
  \centering
  \begin{subfigure}[t]{0.48\columnwidth}
    \centering
    \includegraphics[width=0.78\textwidth]{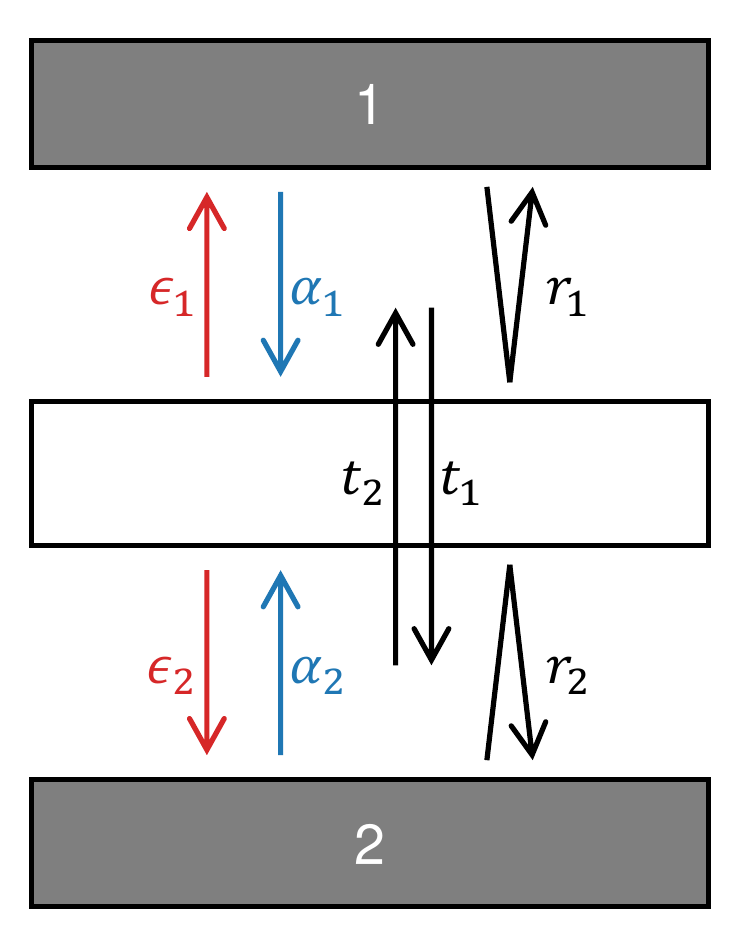}
    \caption{}
    \label{fgr:schematic}
  \end{subfigure}
  \begin{subfigure}[t]{0.48\columnwidth}
    \centering
    \includegraphics[width=\textwidth]{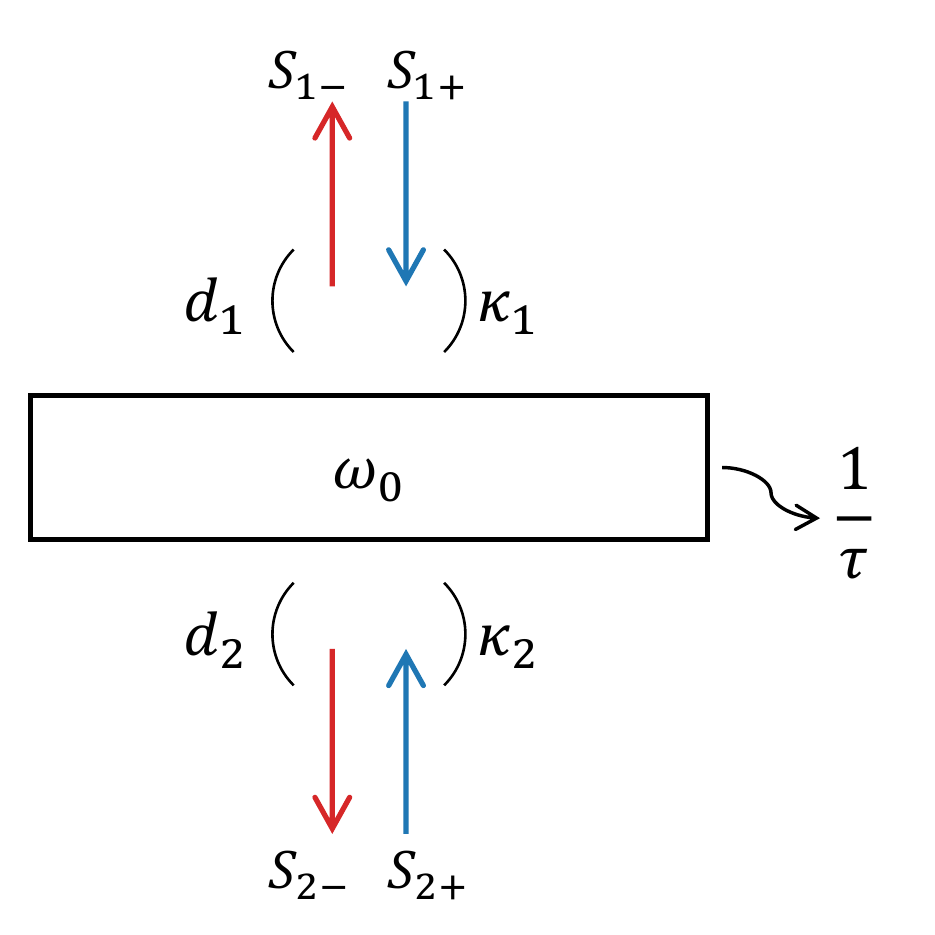}
    \caption{}
    \label{fgr:CMT}
  \end{subfigure}
  \caption{Schematic illustration of the structure. (a)~Energy flow diagram of the considered system with radiation channels above and below an emitter. Each arrow represents a route that an energy flow can take. Blackbodies~1 and 2 are at the same temperature as the emitter in the middle. (b)~Diagram for CMT analysis. The emitter is expressed as a resonator in the middle.}
  \label{fgr:theory}
\end{figure}

  We start by considering the energy flow in the system shown in Fig.~\ref{fgr:schematic}.
  A nonreciprocal emitter (shown as a white box) is located between blackbodies~1 and 2.
  Under the thermal equilibrium, there are several constraints   imposed on the energy flows (shown with arrows  in the figure) in the system which we discuss below.
  Once the constraints are derived, one can apply them to the non-equilibrium situations.  
  In the figure, $\alpha_{1,2}$ represents the absorptivity of the emitter, which accounts for the ratio of energy absorbed by the emitter to the incoming energy for a given illumination direction.
  Emissivity of the emitter $\epsilon_1$ ($\epsilon_2$) accounts for the ratio of energy emitted by the emitter towards blackbody~1 (blackbody~2) to the energy emitted by a blackbody emitter in the same situation.
  Quantities $r_{1,2}$ and $t_{1,2}$ represent the reflectivity and transmissivity, respectively.
  
  Considering first the energy emitted from the blackbody~1 towards the emitter, this energy can take three different routes as shown in Fig.~\ref{fgr:schematic}: it can be absorbed by the emitter ($\alpha_1$), it can be reflected back to the blackbody~1 ($r_1$), or it can transmit through the emitter ($t_1$).
  Thus, the following energy-balance equation has to hold:
\begin{equation}
  \alpha_1+t_1+r_1=1.
  \label{eqn:energywise_1}
\end{equation}
  Similarly considering the energy absorbed by the blackbody~1, this energy consists of three different components: energy emitted from the emitter ($\epsilon_1$), energy that was reflected at the top ($r_1)$, and energy that transmitted through the emitter upwards ($t_2$).
  Therefore, another energy-balance equation holds:
\begin{equation}
  \epsilon_1+t_2+r_1=1.
  \label{eqn:energywise_2}
\end{equation}
  Similar equations can be derived from the blackbody~2 side as well:
\begin{equation}
  \alpha_2+t_2+r_2=1,
  \label{eqn:energywise_3}
\end{equation}
\begin{equation}
  \epsilon_2+t_1+r_2=1.
  \label{eqn:energywise_4}
\end{equation}

  The goal of this work is to design such an emitter which  provides full energy absorption from the top half-space (emitted by blackbody~1) with energy emission exclusively towards the bottom half-space (towards blackbody~2).
  This can be  translated into the conditions of $\alpha_1=1$ and $\epsilon_1=0$ which together with Eq.~\eqref{eqn:energywise_1}--\eqref{eqn:energywise_4} provide us with the requirements:
\begin{equation}
  \epsilon_1=\alpha_2=t_1=r_1=r_2=0,
  \label{eqn:energywise_ideal_0}
\end{equation}
\begin{equation}
  \alpha_1=\epsilon_2=t_2=1.
  \label{eqn:energywise_ideal_1}
\end{equation}
  In other words, under the ideal operation, the emitter will fully absorb energy incident from the top, emit it completely to the bottom, while pass all the energy coming from the bottom without reflection or absorption.
  We can see from Eqs.~\eqref{eqn:energywise_1} and \eqref{eqn:energywise_2} that in this design, the absorptivity-emissivity contrast $|\alpha_1-\epsilon_1|$ is  equal to   the  contrast of transmissivities $|t_1 - t_2|$.
  This is in sharp contrast  with previous designs in Ref.~\citenum{green_time-asymmetric_2012, zhu_near-complete_2014, zhao_near-complete_2019, wang_nonreciprocal_2017, wang_nonreciprocal_2018, pajovic_intrinsic_2020}, where the absorptivity-emissivity contrast arises from the  contrast of reflectivities.

\subsection{Temporal coupled mode theory analysis}
  To achieve the absorptivity-emissivity contrast as discussed above, we consider a resonator system as shown in Fig.~\ref{fgr:CMT}.
  The resonator system couples to two ports on either side of the resonator (one port on each side).
  We apply CMT here, which is valid when the Q-factor of the resonance is sufficiently high, and we assume a single mode resonance.
  According to the CMT formalism, our two-port system can be described by the following equations: \cite{haus_waves_1984, fan_12_2008}
\begin{subequations}
  \begin{align}
    &\frac{d u}{d t} = \left( j\omega_0-\gamma-\frac{1}{\tau} \right) u + \begin{pmatrix} \kappa_1 & \kappa_2 \end{pmatrix} \begin{pmatrix} s_{1+} \\ s_{2+} \end{pmatrix} \\
    &\begin{pmatrix} s_{1-} \\ s_{2-} \end{pmatrix} = C \begin{pmatrix} s_{1+} \\ s_{2+} \end{pmatrix} + \begin{pmatrix} d_1 \\ d_2 \end{pmatrix} u
  \end{align}
  \label{eqn:CMT}
\end{subequations}
  where $u$ is the amplitude of the resonant mode normalized so that $|u|^2$ corresponds to the energy, $\omega_0$ is the resonant angular frequency, $\gamma$ and $1/\tau$ are the decay rates due to radiation and the material loss, respectively. 
  $s_{i+}$ and $s_{i-}$ are incoming and outgoing wave amplitudes ($i=1$ for the top port and $i=2$ for the bottom port), respectively.
  $\kappa_i$ and $d_i$ are coupling coefficients to each ports, $\kappa_i$ for absorption and $d_i$ for emission.
  \[C=\begin{pmatrix} r_{11} & t_{12} \\ t_{21} & r_{22} \end{pmatrix}\]
  is the matrix describing the background scattering process.\cite{fan_temporal_2003}
  If we neglect the material loss for now, i.e. $1/\tau=0$, the system is energy-conserving and the parameters satisfy the following equations:\cite{wang_time-reversal_2018, zhao_connection_2019}
\begin{subequations}
  \begin{align}
    & C^\dagger C = I \label{eqn:CMT_energy_conservation_1}\\
    & C \begin{pmatrix} \kappa^*_1 \\ \kappa^*_2 \end{pmatrix} + \begin{pmatrix} d_1 \\ d_2 \end{pmatrix} = 0 \label{eqn:CMT_energy_conservation_2}\\
    & \left| \kappa_1 \right|^2 + \left| \kappa_2 \right|^2 = \left| d_1 \right|^2 + \left| d_2 \right|^2 = 2\gamma \label{eqn:CMT_energy_conservation_3}.
  \end{align}
  \label{eqn:CMT_energy_conservation}
\end{subequations}
  Note that Eq.~\eqref{eqn:CMT_energy_conservation} only assumes energy conservation and no constraint regarding reciprocity is imposed here. Therefore, this holds for our nonreciprocal structure as well.

  Then, by adding $1/\tau$ back into Eq.~\eqref{eqn:CMT}, we can introduce the scattering matrix $S$ as $ \boldsymbol{s_-} = S \cdot \boldsymbol{s_+}$ which is given by:
\begin{equation}
  S = C + \frac{\boldsymbol{d} \cdot \boldsymbol{\kappa}^\text{T}}{j\left( \omega-\omega_0 \right) + \gamma + 1/\tau}
  \label{eqn:CMT_Smatrix}
\end{equation}
  where $\boldsymbol{s_\pm} = \big(\begin{smallmatrix} s_{1\pm}\\s_{2\pm}\end{smallmatrix}\big)$, $\boldsymbol{d} = \big(\begin{smallmatrix} d_1\\d_2\end{smallmatrix}\big)$, and $\boldsymbol{\kappa} = \big(\begin{smallmatrix} \kappa_ 1\\ \kappa_2\end{smallmatrix}\big)$.
  We note that in order to take account of material loss by simply adding $1/\tau$ back, we have to assume that the influence on other parameters due to this addition is negligible.
  This is a common approximation when applying CMT to describe lossy resonator structures.\cite{piper_total_2014}
  As will be shown below, this approximation works well for the present case.
  
  By comparing the scattering matrix $S$ and the energy flow diagram of Fig.~\ref{fgr:schematic}, we can deduce the following relations: $\left| S_{11} \right|^2 = r_1$, $\left| S_{21} \right|^2 = t_1$, $\left| S_{22} \right|^2 = r_2$, and $\left| S_{12} \right|^2 = t_2$. Therefore, expressions for absorptivities and emissivities can be derived using Eq.~\eqref{eqn:CMT_energy_conservation}:
\begin{subequations}
  \begin{align}
    & \alpha_1 (\omega)= 1 - \left| S_{11} \right|^2 - \left| S_{21} \right|^2  = \frac{2}{\tau} \frac{|\kappa_1|^2}{(\omega-\omega_0)^2+(\gamma+1/\tau)^2}\\
    & \epsilon_1 (\omega)= 1 - \left| S_{11} \right|^2 - \left| S_{12} \right|^2  = \frac{2}{\tau} \frac{|d_1|^2}{(\omega-\omega_0)^2+(\gamma+1/\tau)^2}\\
    & \alpha_2 (\omega)= 1 - \left| S_{22} \right|^2 - \left| S_{12} \right|^2  = \frac{2}{\tau} \frac{|\kappa_2|^2}{(\omega-\omega_0)^2+(\gamma+1/\tau)^2}\\
    & \epsilon_2 (\omega)= 1 - \left| S_{22} \right|^2 - \left| S_{21} \right|^2  = \frac{2}{\tau} \frac{|d_2|^2}{(\omega-\omega_0)^2+(\gamma+1/\tau)^2}.
  \end{align}
  \label{eqn:CMT_AE}
\end{subequations}
  Note that these are Lorentzian functions, and, as will be shown below, they accurately describe the actual spectra  near the resonance frequency for the photonic crystal slab.
  Note also that $\alpha_i$ is proportional to $|\kappa_i|^2$ and $\epsilon_i$ is proportional to $|d_i|^2$ (for $i=1,2$).
  This accords with our initial set-up of CMT, where $\kappa_i$ is the in-coupling coefficient and $d_i$ is the out-coupling coefficient between the port and the resonator.
  
  Since our goal is to maximize the contrast between absorptivity and emissivity,  we now look at the expression for $|\alpha_1-\epsilon_1|$ at resonance frequency $\omega=\omega_0$:
\begin{equation}
  \left| \alpha_1(\omega_0)-\epsilon_1(\omega_0) \right| = \left| \frac{|\kappa_1|^2}{2\gamma} - \frac{|d_1|^2}{2\gamma} \right| \frac{4\gamma\frac{1}{\tau}}{\left( \gamma + \frac{1}{\tau} \right)^2}
  \label{eqn:A-E}
\end{equation}
  It is clear from Eq.~\eqref{eqn:A-E} that maximization of $|\alpha_1(\omega_0)-\epsilon_1(\omega_0)|$ can be achieved by satisfying two separate conditions. The first condition is maximizing the difference between the normalized coupling rates $\frac{|\kappa_1|^2}{2\gamma}$ and $\frac{|d_1|^2}{2\gamma}$. The second condition is the critical coupling condition of $\gamma=1/\tau$, which was also pointed out in Ref.~\citenum{zhu_near-complete_2014}. Thus, a proper design requires simultaneous  maximization of the magnitude difference between the input and output coupling coefficients and  matching of the radiation and absorption losses.

\section{Design}
\begin{figure}
  \includegraphics[width=0.95\columnwidth]{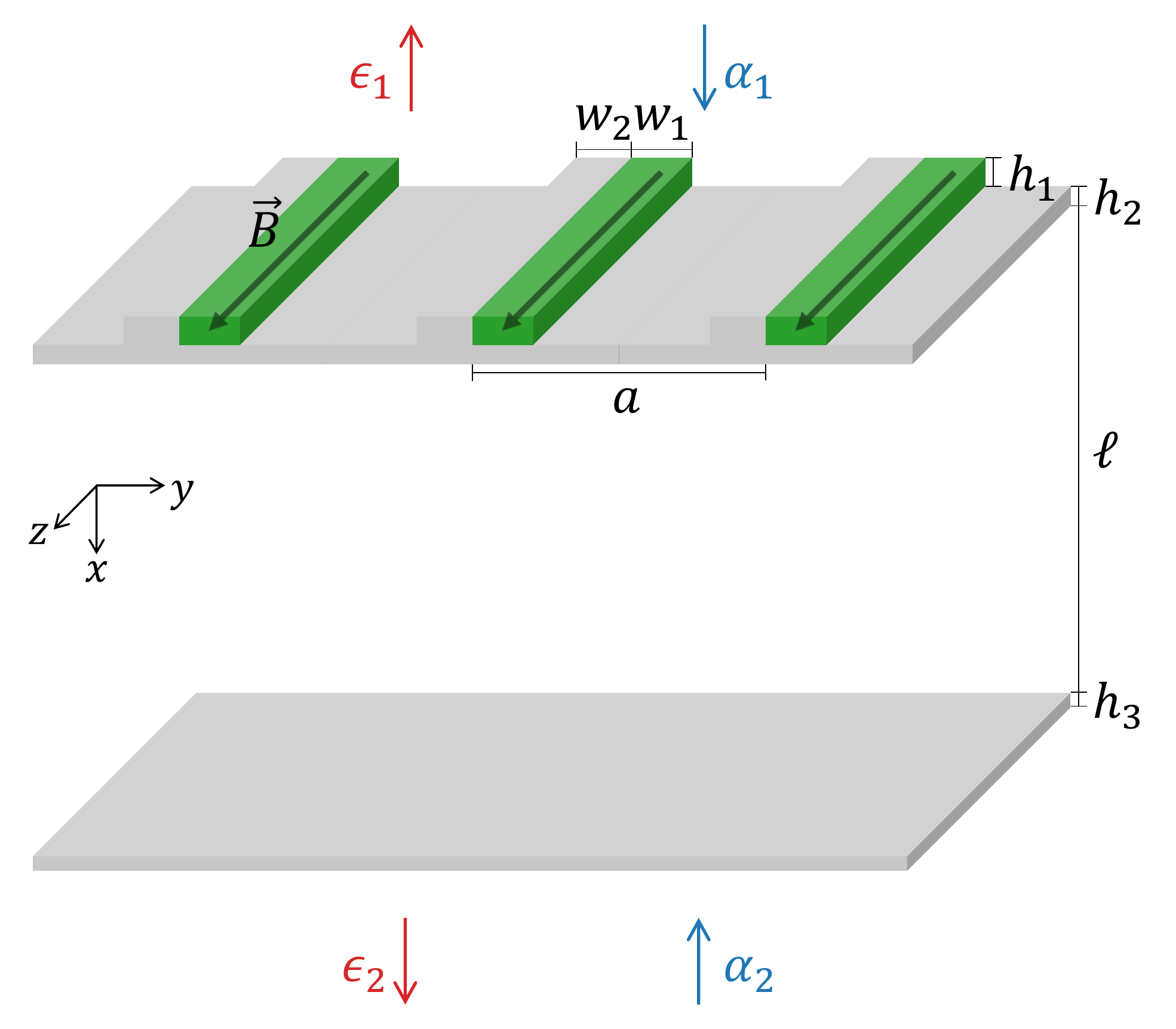}
  \caption{Geometry of the designed nonreciprocal thermal emitter which breaks the Kirchhoff's law. The gray regions represent dielectric, while the green regions represent magneto-optical material with magnetization in the $z$ direction. Two infinitely long bars are located on top of a slab, and another parallel slab is located at a distance $\ell$ away. Wave propagation is along the $\pm x$ direction.}
  \label{fgr:design}
\end{figure}

  Based on the CMT formalism presented in the previous section, we consider an emitter consisting of a photonic crystal structure shown in Fig.~\ref{fgr:design}.
  We place two bars, made of dielectric and magneto-optical materials, on top of a dielectric slab as shown at the top of Fig.~\ref{fgr:design}.
  In addition, we place another parallel dielectric slab at a distance $\ell$ away in the $x$ direction, as shown at the bottom of the figure.
  The entire structure is surrounded by air.
  In this work, we choose the dielectric material to be silicon in order to exploit the strong index contrast between silicon and air. 
  The magnetization direction in the magneto-optical regions is parallel to the $z$ axis, and we express the permittivity of the magneto-optical material by a tensor
\begin{equation}
  \overline{\overline{\varepsilon}}=\begin{bmatrix} \varepsilon_\text{d} & -j\varepsilon_\text{a} & 0 \\ j\varepsilon_\text{a} & \varepsilon_\text{d} & 0 \\ 0&0&\varepsilon_z \end{bmatrix}.
  \label{eqn:permittivity_tensor}
\end{equation}
  Let us decompose the diagonal component into real and imaginary parts so that $\varepsilon_\text{d}=\varepsilon_{\text{dr}}-j\varepsilon_{\text{di}}$.
  The imaginary part accounts for the material loss.
  The off-diagonal component $\varepsilon_\text{a}$ characterizes the strength of the nonreciprocal effect and arises from the magnetization of the material.
  In most situations, the imaginary part of $\varepsilon_\text{a}$ is very small and can be neglected. 
  For incident light, we consider a TM-polarized wave propagating in the $\pm x$ direction with magnetic field in the $z$ direction and electric field in the $xy$ plane.
  The setup therefore is in the Voigt configuration with the wave propagating perpendicular to the direction of magnetization.

  As shown above based on Eqs.~\eqref{eqn:energywise_1} and \eqref{eqn:energywise_2}, maximization of the contrast between $\alpha_1$ and $\epsilon_1$ has the same conditions as maximization of the contrast between the transmission along the two directions.
  Therefore, the design of nonreciprocal emitters is connected with isolator designs that seek to maximize the transmission contrast.
  On the other hand, since our motivation is to construct such a nonreciprocal device in the context of energy harvesting, the device needs to operate in the optical or thermal frequency range for broad-area incident light.
  Thus, many standard designs of isolators are not applicable.
  For example, microwave isolators~\cite{pozar_microwave_2012, kord_microwave_2020} are not appropriate due to low-frequency operation, while optical implementations based on waveguide geometries~\cite{dotsch_applications_2005,williamson_integrated_2020} do not operate with free-space light illuminations. 
  Moreover, those existing optical isolator designs that operate for free-space broad-area incident light at the optical frequencies also have limitations for our purpose.
  The isolators based on the Faraday configuration~\cite{saleh_fundamentals_2019} are generally bulky and additionally require polarizers and antireflective coatings, resulting in the final structure with a large thickness.
  In the Voigt configuration~\cite{zvezdin_modern_1997}, most isolators operate either only in the reflection regime or only for the oblique illuminations~\cite{yu_one-way_2007,asadchy_sub-wavelength_2020}.
  Thus, the challenge for a compact nonreciprocal thermal emitter operating in transmission has not been addressed in previous works.
  
  In order to achieve nonreciprocal emission (or isolation in transmission), the magnetic symmetry group of the structure must exclude some of the symmetry operations.\cite{figotin_nonreciprocal_2001, guo_theoretical_2020}
  For the normal incidence in the $\pm x$ direction, the representing examples of the symmetries that simultaneously need to be broken are $R$, $I$, $m_x$, $m_y'$, $m_z'$, $n_x'$, $p_y$, and $p_z$.
  Here, $R$ represents the time-reversal symmetry operation, and $I$ represents the inversion symmetry operation.
  $m_j$ represents the mirror symmetry with respect to the plane perpendicular to the $j$ axis, while $n_j$ and $p_j$ represent the $n$-fold and $p$-fold rotation symmetry, respectively, about the $j$ axis, where $n$ is any natural number higher than 1 and $p$ is any even natural number.
  Index $j$ can be any one of $x$, $y$, and $z$, while $'$ means an additional time-reversal operation to the preceding operation (e.g. $m_j'=m_j\odot R$).
  Considering only the array of bars in Fig.~\ref{fgr:design} (excluding both continuous slabs), we can see that their symmetry group includes only the symmetry operations $m_x'$, $m_z$, and $2_y'$, which does not preclude the nonreciprocal emission.
  Therefore, the array of bars consisting of dielectric and magneto-optical materials (with the given magnetization direction) by itself can exhibit the desired nonreciprocal property.
  However, the nonreciprocal effect in such a system is generally weak due to its low Q-factor.
  Hence, we add two continuous slabs: the upper one for creating a strong guided-mode resonance response~\cite{fan_temporal_2003} and the bottom one, placed sufficiently far apart, for independent tuning of the background scattering matrix $C$ in our CMT model without affecting the total linewidth and the resonant frequency.
  The bottom slab provides us with an additional degree of freedom useful for simplifying the design.
  The only symmetry present in the final system in Fig.~\ref{fgr:design} is $m_z$, which satisfies the symmetry requirement as discussed above.
  
  In the structure of Fig.~\ref{fgr:design}, for TM-polarized light normally incident onto the structure, the part of the scattered field that arises from the magneto-optical materials has the same pattern independent whether the light is incident from the top or the bottom.
  Moreover, the scattered field above or below the structure exhibits a 180 degree phase difference.
  The dielectric structures surrounding the magneto-optical materials have the right symmetry to ensure that such scattered field can radiate into far field.
  Consequently, there is constructive or destructive interference depending on the direction of incident light, which results in the transmission contrast.

\section{Results and discussion}
\subsection{Optimization for ideal performance}
  Here, we explain the optimization of our design with the guidance of CMT model developed above.
  First, to illustrate the general principles, we conduct the optimization without restricting ourselves to any specific materials.
  Then, in the following section, we discuss more about the material constraints.

  We conduct full-wave simulations of the emitter shown in Fig.~\ref{fgr:design} using a 2D finite-element frequency domain solver (RF module in COMSOL Multiphysics).
  Periodic boundary condition is applied in the $xy$ plane.
  We compute the spectra of reflectivities and transmissivities as we illuminate the nonreciprocal structure by TM-polarized light along either the $+x$ or $-x$ direction (see Fig.~\ref{fgr:design}).
  We calculate using Eq.~\eqref{eqn:CMT_AE} the spectra of absorptivities and emissivities from the reflectivities and transmissivities.
 
  In the simulations, we optimize the structural parameters of the emitter subject to the following material properties.
  For the dielectric regions, we set the dielectric constant to $\varepsilon_\text{s}=12$, which approximately corresponds to that of silicon in the infrared frequency range.
  For the magneto-optical material, we choose the real part of the diagonal relative permittivity to be $\varepsilon_\text{dr}=6.25$.
  This corresponds to that of bismuth iron garnet (BIG), also in the infrared frequency range, which has been a widely used material for magneto-optical applications in several preceding studies.\cite{adachi_epitaxial_2000, tepper_pulsed_2003, wang_optical_2005, yu_one-way_2007}
  Under these constraints, we first tune the geometrical parameters, i.e. $w_1$, $w_2$, $h_1$, $h_2$, $h_3$, and $\ell$ shown in Fig.~\ref{fgr:design}, as well as the off-diagonal relative permittivity $\varepsilon_\text{a}$, so that we can maximize $\left| \frac{|\kappa_1|^2}{2\gamma} - \frac{|d_1|^2}{2\gamma} \right|$ in Eq.~\eqref{eqn:A-E}.
  Here, we choose the periodicity of our structure sub-wavelength so that no diffraction harmonics are excited.
  In this process, we fit the absorptivity and emissivity spectra obtained from the simulation to the expressions derived from our CMT model (Eq.~\eqref{eqn:CMT_AE}), to determine $\kappa_i$, $d_i$, and $\gamma$.
  Once the contrast close to unity is achieved, we then tune the imaginary part of the diagonal relative permittivity $\varepsilon_\text{di}$ to satisfy the critical coupling condition. 
  Since $1/\tau$, which is also determined from the fitting, can be controlled independently from other CMT parameters by varying $\varepsilon_\text{di}$, we can safely tune the material loss at the end without affecting the parameters determined in the previous step.
  
  As a result of optimization, $\varepsilon_\text{a}=0.19$ and $\varepsilon_\text{di}=0.0015$ are chosen.
  These values are approximately 3~times larger and 60~times smaller than the actual values for BIG, respectively.
  The optimized geometric parameters read as: $w_1=0.21a$, $w_2=0.19a$, $h_1=0.097a$, $h_2=0.065a$, $h_3=0.05a$, and $\ell=1.67a$, where $a$ is the periodicity.
  All the geometric parameters  are illustrated in Fig.~\ref{fgr:design} with the correct scaling.

\begin{figure}
  \centering
  \begin{subfigure}[t]{0.48\columnwidth}
    \centering
    \includegraphics[width=\textwidth]{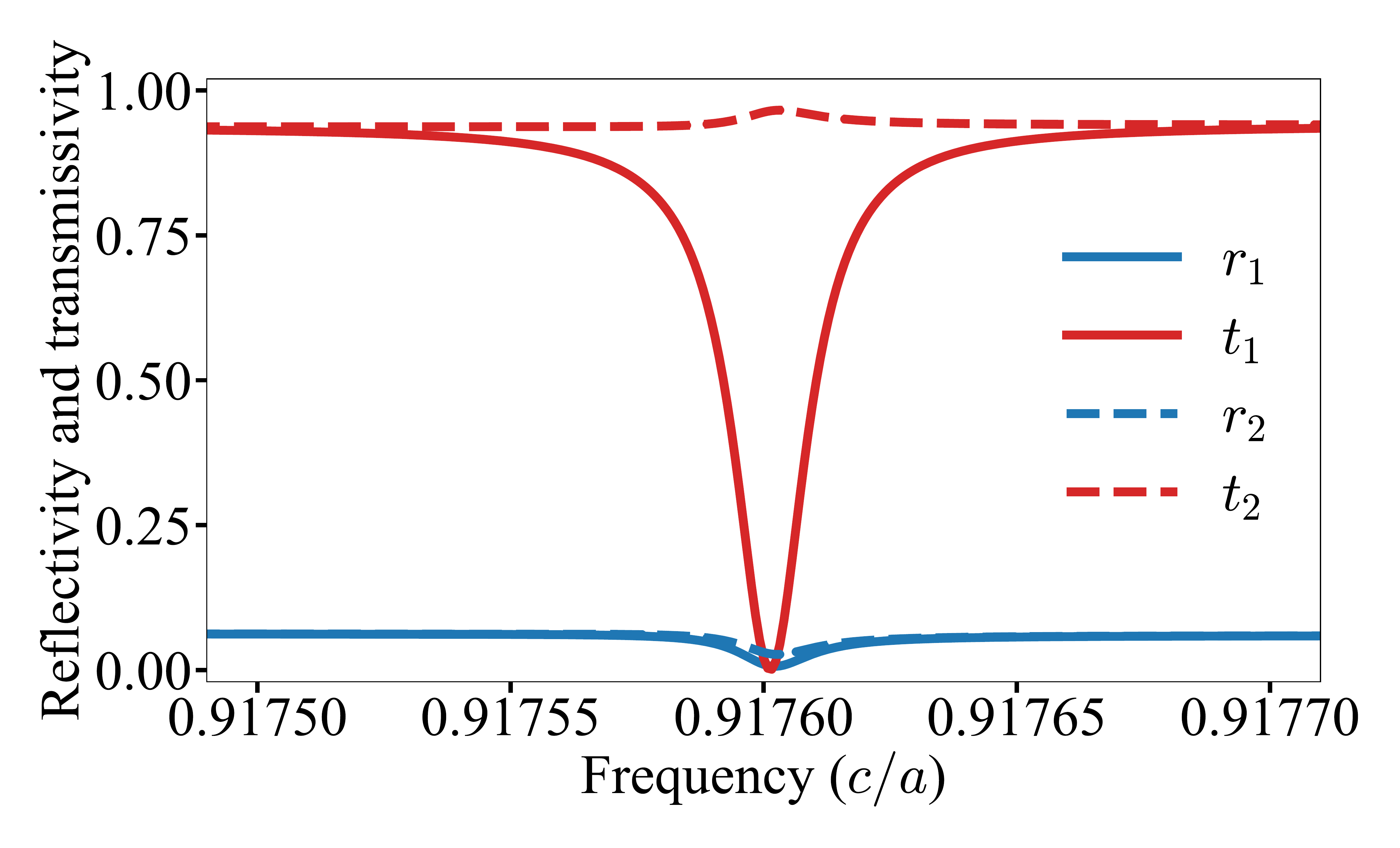}
    \caption{}
    \label{fgr:RT_ideal}
  \end{subfigure}
  \begin{subfigure}[t]{0.48\columnwidth}
    \centering
    \includegraphics[width=\textwidth]{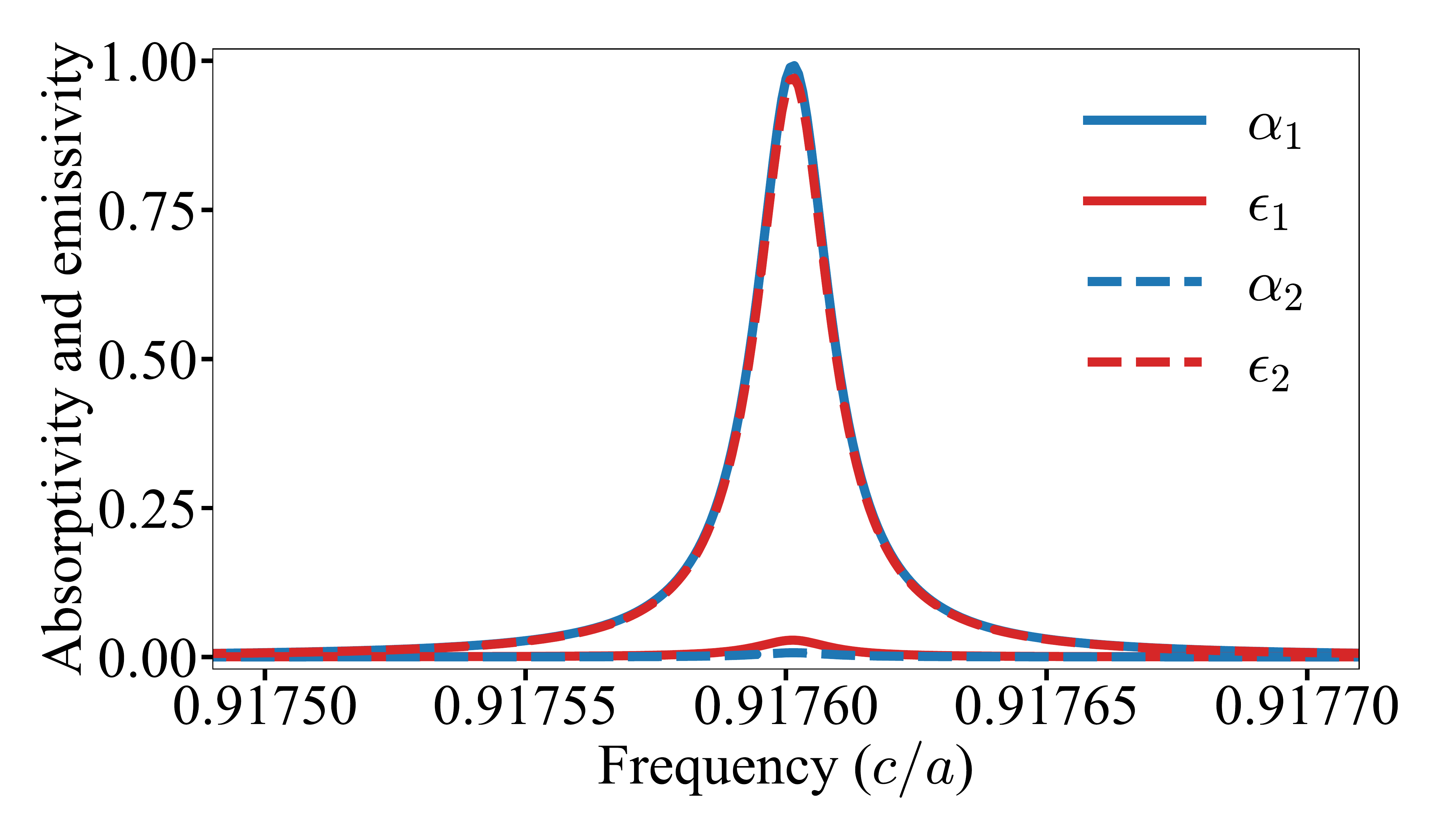}
    \caption{}
    \label{fgr:AE_ideal}
  \end{subfigure}
  \caption{Response functions of the optimized nonreciprocal emitter with nearly ideal performance. The shown results are for the TM-polarized light. (a)~Reflectivities and transmissivities. (b)~Absorptivities and emissivities.}
  \label{fgr:result_ideal}
\end{figure}
  
  Fig.~\ref{fgr:result_ideal} shows the response functions for the optimized emitter.  
  As one can see, a nearly complete contrast between the absorptivity and emissivity is achieved, i.e. $|\alpha_1(\omega_0)-\epsilon_1(\omega_0)|=0.965$.
  At the resonance the reflectivities are negligible, so that absorptivity-emissivity contrast comes from the transmissivity contrast.
  We note that our design procedure is general and can be applied to arbitrary frequencies by tuning the material and structural properties.
  
\begin{figure}
  \includegraphics[width=0.95\columnwidth]{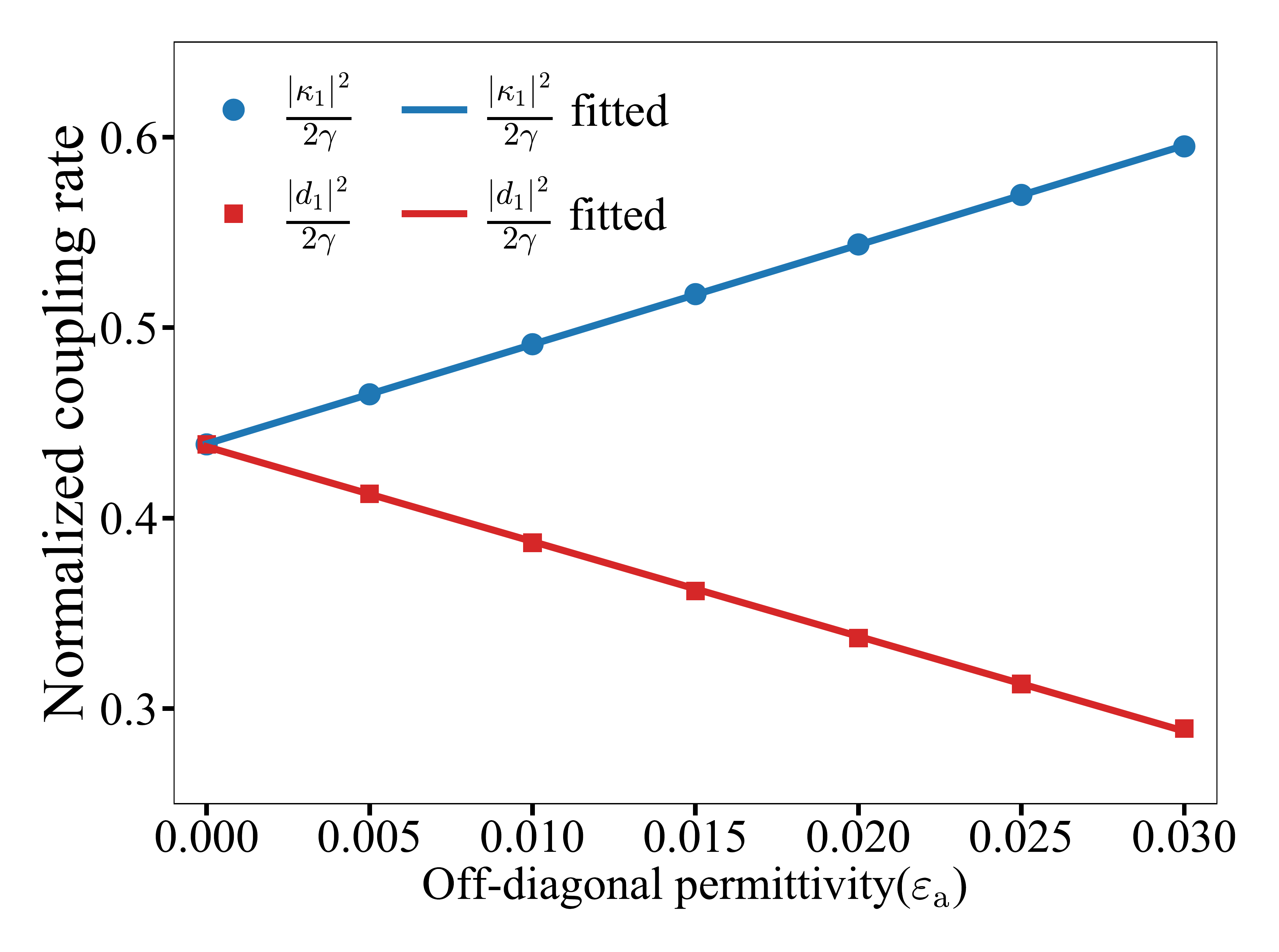}
  \caption{The normalized coupling rates versus the off-diagonal permittivity. For small values of $\varepsilon_\text{a}$, the splitting of the input and output coupling rates is linear. The geometrical parameters of the structure are determined from the optimization process.}
  \label{fgr:dk_splitting}
\end{figure}
  
  As we can see from Eq.~\eqref{eqn:A-E}, a key consideration for achieving high contrast between the absorptivity and emissivity is to achieve a large difference between the in-coupling coefficient $\kappa_1$ and out-coupling coefficient $d_1$.
  Therefore, below we examine the influence of the magneto-optical effect on such coupling constants, by considering the dependency of the normalized coupling rates $\frac{|\kappa_1|^2}{2\gamma}$ and $\frac{|d_1|^2}{2\gamma}$ as functions of $\varepsilon_\text{a}$ (Fig.~\ref{fgr:dk_splitting}).
  Note that we tune $\varepsilon_\text{a}$ in a range where it is sufficiently small to be considered as a perturbation, while maintaining the other parameters as they were determined from the optimization process.
  
  When $\varepsilon_\text{a}=0$, the input and output coupling rates exactly match, and this corresponds to the reciprocal case without magnetization.
  As $\varepsilon_\text{a}$ increases, the two coupling rates split and the relation between this splitting and $\varepsilon_\text{a}$ appears to be linear. 
  This linearity can be explained by applying first-order perturbation theory~\cite{joannopoulos_photonic_2008} to the relation between the coupling rates and the Hamiltonian of the system, and by considering a time-reversal conjugate system.
  The out-coupling rate from a resonator to an adjacent port has the following relation to Hamiltonian $V$:~\cite{fan_theoretical_1999}
\begin{equation}
  |d_1|^2 \propto \overline{\left|\left<c|V|q\right>\right|^2} g(\omega)
  \label{eqn:Hamiltonian}
\end{equation}
  where $\left| q \right>$ and $\left| c \right>$ represent a state in the port and the resonator, respectively, $V$ is the Hamiltonian that describes the coupling between the port and the resonator, the overline denotes the average over the states in the port, and $g$ is the density of states in the port.
  Then, if we consider small $\varepsilon_\text{a}$ as a perturbation applied to the original system, the following approximation holds:
\begin{equation}
\begin{aligned}
  &\Delta |d_1|^2 \propto \overline{\left| \left<c|V+\delta V|q\right> \right|^2} - \overline{\left| \left<c|V|q\right> \right|^2} \\
  &\approx \overline{\left<q|V|c\right> \left<c|\delta V|q\right> + \left<c|V|q\right> \left<q|\delta V|c\right>}
  \label{eqn:perturbation}
\end{aligned}
\end{equation}
  where $\delta V$ is the perturbation of the Hamiltonian coming from the addition of $\varepsilon_\text{a}$.
  Therefore, when $\varepsilon_\text{a}$ is sufficiently small, change in the out-coupling rate $\Delta |d_1|^2$ is linearly proportional to $\varepsilon_\text{a}$ due to Eqs.~\eqref{eqn:Hamiltonian} and \eqref{eqn:perturbation} since the Hamiltonian elements $\left<c|\delta V|q\right>$ and $\left<q|\delta V|c\right>$ are linear with respect to $\varepsilon_\text{a}$.
  Meanwhile, the in-coupling coefficient $\kappa_1$ in the emitter corresponds to the out-coupling coefficient $\tilde{d}_1$ in its time-reversal conjugate system~\cite{zhao_connection_2019} (tilde symbol denotes the conjugation operator) since these coefficients can be determined in the corresponding lossless system as we discussed in the CMT analysis section.
  Then, since the conjugate emitter has the off-diagonal permittivity $\varepsilon_\text{a}$ with the opposite sign, the slope of $\Delta |\tilde{d}_1 |^2=\Delta |\kappa_1|^2$ will be opposite to that of $\Delta |d_1|^2$.
  This behavior leads to the linear and opposite splitting of the coupling rates $|d_1|^2$ and $|\kappa_1|^2$.
  On the other hand, if we consider $\tilde{\gamma}$ in the conjugate system, due to Eq.~\eqref{eqn:CMT_energy_conservation_3} and the relationship between the coupling coefficients of original and conjugate system ($\tilde{d}_i=\kappa_i$ and $\tilde{\kappa}_i=d_i$ for $i=1,2$), $\tilde{\gamma}$ has to be equal to $\gamma$ of the original system.
  As a result, $\gamma$ is an even function with respect to $\varepsilon_\text{a}$ and constant in the first-order approximation.
  Thus, the normalized coupling rates $\frac{|d_1|^2}{2\gamma}$ and $\frac{|\kappa_1|^2}{2\gamma}$ shown in Fig.~\ref{fgr:dk_splitting} also vary linearly with respect to $\varepsilon_\text{a}$, with different slope signs.
  Note that this linear and monotonic increase of splitting is limited to the regime where $\varepsilon_\text{a}$ is sufficiently small, and similar behavior is not assured when the first order approximation does not hold anymore.

\subsection{Emitter realization based on Weyl semimetal}
\begin{figure}
  \centering
  \begin{subfigure}[t]{0.48\columnwidth}
    \centering
    \includegraphics[width=\textwidth]{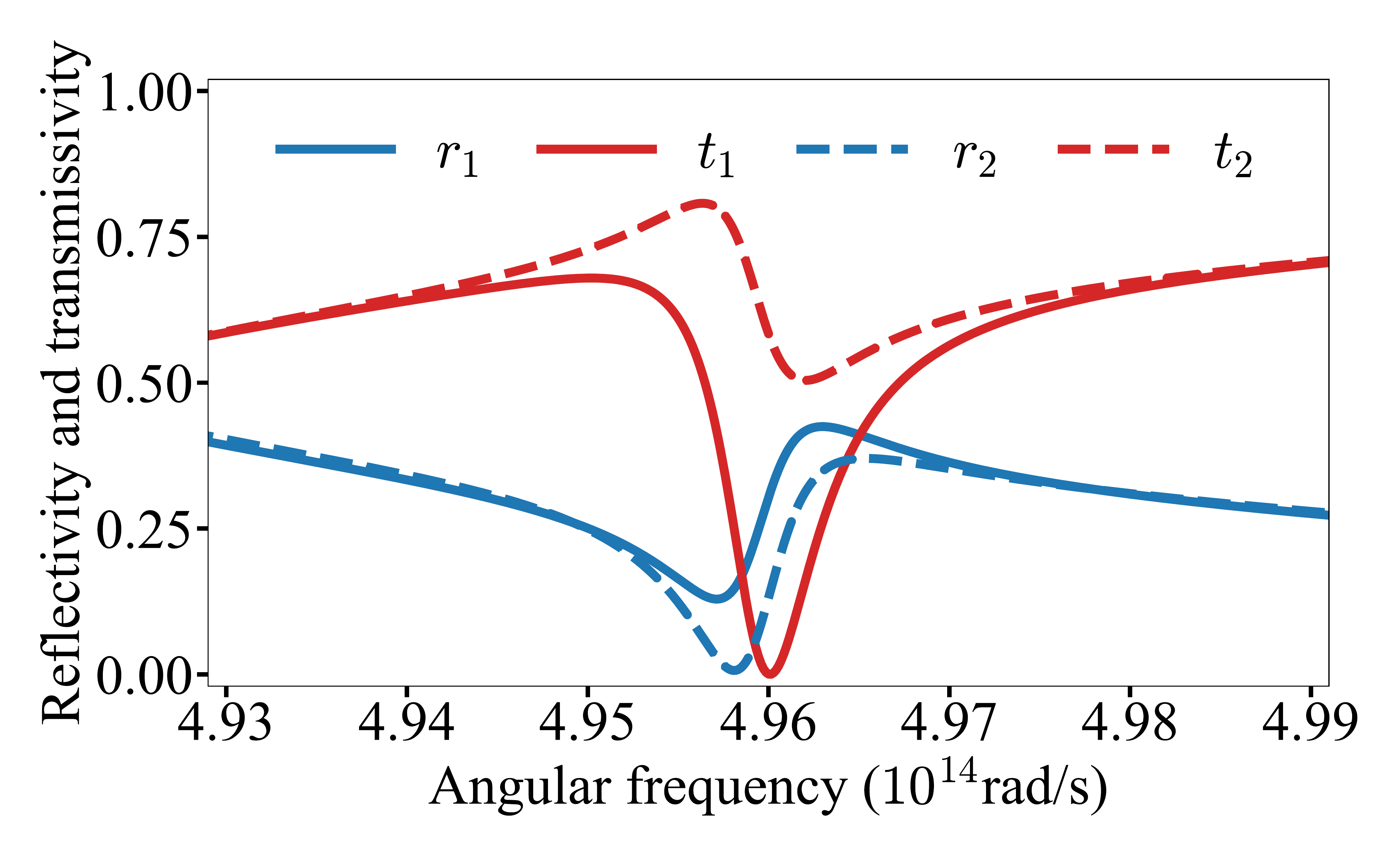}
    \caption{}
    \label{fgr:RT_Weyl}
  \end{subfigure}
  \begin{subfigure}[t]{0.48\columnwidth}
    \centering
    \includegraphics[width=\textwidth]{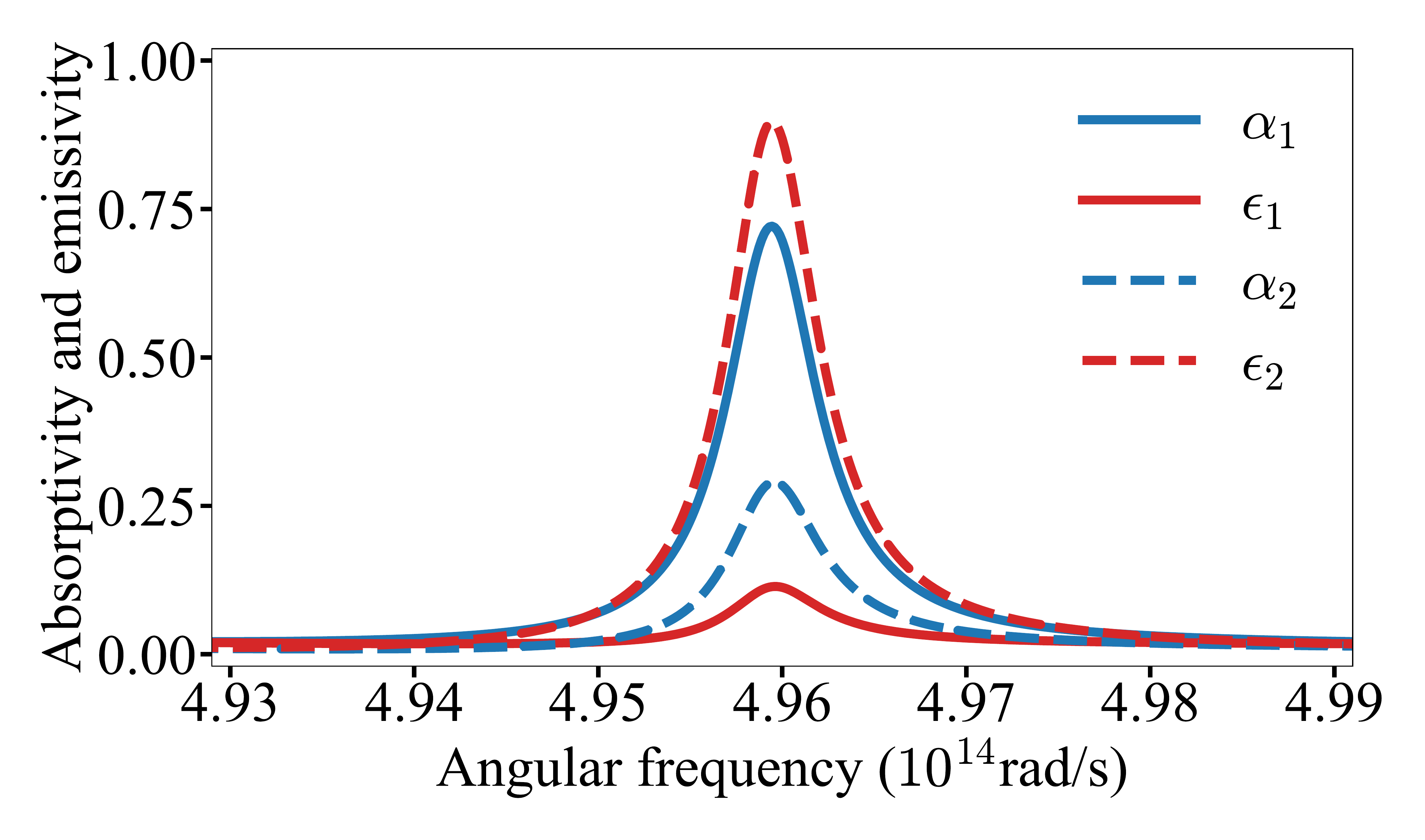}
    \caption{}
    \label{fgr:AE_Weyl}
  \end{subfigure}
  \caption{Response functions of the optimized emitter consisting of Weyl semimetal and silicon. The shown results are for the TM-polarized light. (a)~Reflectivities and transmissivities. (b)~Absorptivities and emissivities.}
  \label{fgr:result_Weyl}
\end{figure}

  Although the results shown above demonstrate nearly perfect nonreciprocal emitter performance, finding magneto-optical material whose permittivity is equal to the optimized one can be a  nontrivial task.
  For example, BIG with realistic losses in the diagonal permittivity and realistic $\varepsilon_\text{a}$ will not provide a high emissivity-absorptivity contrast.
  As a rule of thumb, the high contrast requires sufficiently small $\varepsilon_\text{di}$ and sufficiently large $\varepsilon_\text{a}$ of the magneto-optical material.
  Here, we propose to use a magnetic Weyl semimetal of EuCd\textsubscript{2}As\textsubscript{2} as one candidate of the magneto-optical material with such desired properties.
  
  We use the permittivity dispersion data from Refs.~\citenum{zhao_axion-field-enabled_2020, asadchy_sub-wavelength_2020, guo_radiative_2020}.
  While this material is relatively lossy, it has an extremely high off-diagonal permittivity, which makes its overall performance good.
  Since $\varepsilon_\text{a}$ is large enough at most frequencies, we consider the operating frequency that minimizes the $\varepsilon_\text{di}/\varepsilon_\text{dr}$ value.
  At frequency $\omega_0=4.96\times10^{14}\si{\radian/\second}$, the permittivity values for the Weyl semimetal are $\varepsilon_\text{d}=3.47-j0.0733$ and $\varepsilon_\text{a}=-2.39$.
  Since the operating frequency is fixed now, we set $\varepsilon_\text{s}$ to 11.7, which corresponds to the dielectric constant of silicon at this frequency.
  Next, using the approach described above we optimize the geometry of the Weyl emitter at the resonance frequency $\omega_0$.
  The optimized geometric parameters read as: $a=1.61\si{\micro\metre}$, $w_1=400\si{\nano\metre}$, $w_2=220\si{\nano\metre}$, $h_1=830\si{\nano\metre}$, $h_2=700\si{\nano\metre}$, $h_3=80\si{\nano\metre}$, and $\ell=2.13\si{\micro\metre}$.
  We note that the bandwidth of the resonance is about $10^{12}\si{\radian/\second}$, which is narrow enough to neglect the impact of dispersion, and we assume constant permittivity values throughout this resonance regime.
  
  As can be seen from Fig.~\ref{fgr:RT_Weyl}, the  asymmetric Fano line shape of the guided-mode resonance appears in the reflectivity and transmissivity spectra.
  At this resonance, the absorptivity-emissivity contrast reaches $|\alpha_1(\omega_0)-\epsilon_1(\omega_0)|\approx 0.6$ (see Fig.~\ref{fgr:AE_Weyl}).
  Therefore, relatively high contrast in absorptivity and emissivity can be achieved with realistic material systems.
  Importantly, the operating  frequency of the nonreciprocal emitter can be shifted by changing the Fermi level of the Weyl semimetal    \cite{zhao_axion-field-enabled_2020,asadchy_sub-wavelength_2020}.  
  
  Note that when considering the electromagnetic properties of a Weyl semimetal, it is important to take account of the Fermi-arc surface states that can affect the boundary conditions for the electromagnetic waves.\cite{chen_optical_2019}
  However, in our simulation, the magnitude of the tangential electric and magnetic fields along the boundary of the Weyl semimetal region turns out to meet the condition of neglecting the Fermi-arcs ($\frac{e^2}{h}\frac{|E_t|}{|H_z|} \ll 1$ where $E_t$ is the tangential electric field and $H_z$ is the magnetic field in $z$ direction) proposed in Ref.~\citenum{asadchy_sub-wavelength_2020}.
  Thus, we ignore this effect and consider the conventional boundary conditions.

\section{Conclusion}
  In conclusion, we show that it is possible to violate the Kirchhoff's law of radiation in a semitransparent structure.
  We develop a CMT framework to guide the design of a nonreciprocal thermal emitter that fully absorbs normally incident energy from a given direction with zero backward but unity forward emissivity.
  Unlike previous reflection-based designs, we suppress reflections and acquire absorptivity-emissivity contrast from difference in transmissivities for each illumination direction.
  Exploiting the guided-mode resonance, we obtain nearly ideal performance for the emitter which has thickness of 1.7 times of the operating wavelength.
  In addition, we demonstrate that the normalized coupling rates in the CMT model are linear functions of the magneto-optical off-diagonal permittivity in the small perturbation limit.
  We also show that, taking into account limited permittivity spectrum of natural materials, it is still possible to reach a significant contrast between the emissivity and absorptivity using recently discovered magnetic Weyl semimetal of EuCd\textsubscript{2}As\textsubscript{2}.
  When applied to multi-junction solar cells, our semitransparent structure will allow the reuse of light energy that would be wasted otherwise.
  Furthermore, directional transmission in our structure coincides with the function of nonreciprocal  isolators.
  Thus, our work may serve as a stepping stone for further utilization of nonreciprocal systems in novel energy harvesting devices and optical isolators.

\begin{acknowledgement}
This work was supported by the Department of Energy Photonics at Thermodynamic Limits Energy Frontier Research Center under Grant DE-SC0019140 (theory and design), and by the Defense Advanced Research Projects Agency Grant No. HR00111820046 (Numerical Simulations).
\end{acknowledgement}

% \begin{suppinfo}

% A listing of the contents of each file supplied as Supporting Information
% should be included. For instructions on what should be included in the
% Supporting Information as well as how to prepare this material for
% publications, refer to the journal's Instructions for Authors.

% The following files are available free of charge.
% \begin{itemize}
%   \item Filename: brief description
%   \item Filename: brief description
% \end{itemize}

% \end{suppinfo}

\bibliography{achemso-demo}

\end{document}